\documentclass[twocolumn,preprintnumbers,amsmath,amssymb]{revtex4}
\usepackage{graphicx}
\usepackage{dcolumn}
\usepackage{bm}

\begin{document}
\title{Polaron tunneling dynamics in the DNA molecule}
\author{Julia A. Berashevich$^*$, Vadim Apalkov$^\dagger$, and 
Tapash Chakraborty$^*$}
\affiliation{$^*$Department of Physics and Astronomy, University of
Manitoba, Winnipeg, MB R3T 2N2, Canada \\
$^\dagger$Department of Physics and Astronomy, Georgia State
University, Atlanta, Georgia 30303, USA}

\begin{abstract}
The formation of polaron and its migration in a DNA chain are 
studied within a semiclassical Peyrard-Bishop-Holstein 
polaron model. Comparing the energetics of the polaron system 
found from the quantum chemical and semiclassical calculations,
we extract the charge-phonon coupling constant for poly DNA 
sequences. The coupling constant is found to be larger for the 
G-C than for the A-T pairs. With this coupling constant 
we study tunneling in the DNA molecule. The rates and the nature 
of tunneling have strong dependence on the DNA sequence. By changing 
the trap positions in the molecular bridge the tunneling 
rate can be varied up to seven orders of magnitude. 
\end{abstract}
\maketitle

The discovery of conductance in DNA has attracted many researchers 
to investigate the transport properties \cite{review,ratner,%
berlin,yoo,kasumov,kawai,porath,conw,jortner} of DNA. For the 
(G-C)(A-T)$_N$(G-C)$_3$ DNA sequences the mechanism of charge transfer 
is more or less clear \cite{review,ratner,berlin,jortner} and is 
described by the competiton between tunneling and hopping 
transfer. But for the poly and mixed DNA sequences the experimental 
data observed by different groups are often contradictory. In some 
experiments a high conductivity was obtained \cite{yoo, kasumov}, 
while in others the conductivity was rather low \cite{kawai,porath}. 
In the works devoted to simulation of charge transfer in the DNA 
molecule within the tight-binding Hamiltonian \cite{macia,wang} or the system of 
kinetic equations \cite{jortner,berlin}, no explanation of this 
phenomena was found. Within these models the charge transfer integral 
between the nearest base pairs and the energy gap between the states 
were the key parameters. At the same time the models did not take into
account consistently the effect of geometry fluctuations or phonons on
the charge transfer processes in the DNA molecule. This is despite the 
fact that already in 1956 Marcus pointed out \cite{mark1} that 
geometry fluctuations can activate and strongly affect an electronic 
transition between two states. Therefore, the polaron model \cite{Bishop1,conw,Bishop2,yoo}, 
which takes into account all these parameters should be invoked for an 
adequate investigation of the charge transfer in DNA.

In this paper, we demonstrate the possibility to design an artificial 
DNA molecule with semiconductor or insulating behaviors simply by 
placing a trap at the correct points. The study is based on the 
analysis of charge transfer from the donor to the acceptor through a 
molecular bridge composed of the potential barriers (A-T pairs) and 
the charge traps (G-C pairs). We show that in the mixed DNA molecules 
the transfer rate of the charge strongly depends on the sequences, 
i.e., on the positions of the traps between the donor and the acceptor. 
This position determines the relation between the rate of charge 
trapping and the rate of charge escape from the trap. Depending on 
this ratio the tunneling between the donor and the acceptor 
can be described either as a sequential tunneling or coherent
tunneling through a trap. This difference strongly affects the final 
rate of charge transfer. 
We show that the time of polaron tunneling can be changed by 
$10^7$ times by changing the position of the trap. The slower 
transfer occurs when the polaron trapping rate is much slower 
than the escape rate. On the other hand, the fastest polaron 
transfer occurs when the trapping and the escape rates are equal. 
In this case the tunneling has the coherent nature and the polaron  
only partially occupies the trap. 

We focus here on the DNA structures where the guanine 
and the adenine are stacked in one DNA strand.  In this case, 
there is only one degree of freedom for the charge transfer  
-- the longitudinal one-dimensional charge tunneling through a 
single strand. The polaron tunneling is described by the system 
of equations within the Peyrard-Bishop-Holstein 
(PBH) model \cite{Bishop1,Bishop2}, where the charge motion is treated 
quantum-mechanically while the polaron tunneling classically. 
The Schr\"odinger equation describing the dynamics of the charge within
the DNA chain with $n$ sites is determined by the Hamiltonian $\cal H$, which 
includes the tight-binding and the charge-lattice interaction 
terms. The Schr\"odinger equation has the form 
\begin{equation}
i\hbar\frac{d\Psi_i}{dt}=-V_{i-1,i}\Psi_{i-1}-V_{i,i+1}\Psi_{i+1}+
\chi_i y_i\Psi_i-\epsilon_i\Psi_i ,
\label{eq:one}
\end{equation}
where $\Psi_i$ is the probability amplitude for the charge to be  
on the $i$-th base pair, $V_{i-1,i} (V_{i,i+1})$ is the 
transfer integral between the nearest base pairs, 
$\chi_i$ is the charge-vibrational coupling 
constant for the $i$-th site, $\epsilon_i$ is the on-site energy, 
$y_i$ determines the stretching at the $i$-th site, i.e. 
the displacement of the atomic structure. 
The motion of the stretching displacement $y_i$ 
is described by the Newton equation as \cite{Bishop2}
\begin{eqnarray}
m\frac{d^2y_i}{dt^2} &=&-V^{\prime}_M(y_i)-W^{\prime}(y_i,y_{i-1})
-W^{\prime}(y_{i+1},y_i) \nonumber\\
&-& \chi|\Psi_i|^2-m\gamma\frac{dy_i}{dt}
\label{eq:two}
\end{eqnarray}
where $m$ is the base pair mass, $\gamma$ is the friction 
parameter ($\gamma=1$ ps$^{-1}$ \cite{Bishop1}), 
$V_M(y_i)$ is the vibronic potential of the 
deformation energy of the hydrogen bonds in the base pair 
taking into acount the repulsive interactions of the phosphates
and $W(y_i,y_{i-1})$ is the nearest-neighbor potential of 
interactions of the stacked base pairs \cite{Bishop2}.
The parameters for the deformation potential and the interaction 
potential are taken from Ref.~\cite{Bishop2}. 

\begin{table}
\caption{\label{tab:table1} The values of 
$\frac12\lambda_N$ and $\chi_i$ for the (G-C)$_N$ and 
the (A-T)$_N$ complexes 
($V_{i-1,i}=V_{i,i+1}=V_i$=0.1 eV).}
\begin{ruledtabular}
\begin{tabular}{ccc}
& $\frac12\lambda_N$ (eV) \footnotemark[1] 
&$\chi_i$ (eV/\AA) \\
\hline
(G-C)$_{N=1}$ & 0.360 & 1.60 \\
(G-C)$_{N=2}$ & 0.310 & 1.15 \\
(G-C)$_{N=3}$ & 0.265 & 0.90 \\
(G-C)$_{N=4}$ & 0.225 & 0.60 \\
(A-T)$_{N=1}$ & 0.185 & 1.05 \\
(A-T)$_{N=2}$ & 0.165 & 0.53 \\
(A-T)$_{N=3}$ & 0.140 & 0.40 \\
(A-T)$_{N=4}$ & 0.125 & 0.30 \\
\end{tabular}
\end{ruledtabular}
\footnotetext[1]{Simulation results are from Ref.~\cite{berash}}
\end{table}

The probability amplitude and the position of the stretching 
displacement in time are evaluated from the self-consistent 
solution of the time-dependent Schr\"odinger and Newton 
equations. Here the Schr\"odinger equation (1) is presented as  
\begin{equation}
\left[1+\frac{i}{2\hbar}{\cal H}\Delta t\right] 
\Psi(t+\Delta t)
=\left[1-\frac{i}{2\hbar}{\cal H}\Delta t\right]\Psi(t)
\label{eq:five}
\end{equation}
where the Hamiltonian $\cal H$ is a function of the
stretching displacements $y_i(t)$. The three-point difference 
scheme for the first order time derivation of the stretching 
parameter has been used in  Eq.~(\ref{eq:two}).
For the stacionary solution of Eqs. (\ref{eq:one}-\ref{eq:two}), 
the initial occupation probability 
is assumed to be close to unity on the donor site. 
This solution is taken as the initial state ($t$=0) for studying the polaron dynamics.

At first we analyze the equilibrium stationary 
polaronic states within a finite region of the DNA chain.  
The charge-vibration coupling constant $\chi_i$ is the main 
parameter regulating the stretching of the polaron 
to the nearest sites \cite{Bishop2} and the magnitude of 
$y_{i}$. The value of $\chi_i$ 
depends on the geometries of the DNA sites participating in 
the formation of the polaron. 
The shift of the state energy due to the polaron
occupation $\chi_i y_i$ in the absence of the DNA-solvent interactions can be 
described by the inner-sphere reorganization energy 
($\approx 0.5\lambda_i$ in Ref.~\cite{berash})
\begin{equation}
\frac12\lambda_N \approx \sum_{i=1}^N \chi_i y_i,
\end{equation}
where $N$ is number of sites occupied by the polaron.
Recently, the exponential decrease of 
the inner-sphere reorganization energy with the 
elongation of the DNA chain was found within the quantum 
chemical calculations for the (G-C)$_N$ and the (A-T)$_N$ 
chains \cite{berash}. The geometry relaxation was found 
to have a maximum at the center of the polaron, which agrees 
with the results obtained within the PBH model \cite{Bishop2}. 

The reorganization energy $\lambda_N$ found in 
Ref.~\cite{berash} are shown in Table~I 
for the (G-C)$_N$ and the (A-T)$_N$ chains. 
From these data we can estimate 
the coupling constants $\chi_i$ for different systems.
The corresponding results for $\chi_i$ are shown 
in Table~I. We have found that the coupling constant 
is smaller for the A-T base pair than for the G-C pair. 
We also determine the tendency in the dependence of the 
coupling constant on the size of the complexes;  
the coupling constant decreses with increasing sizes of the
(G-C)$_N$ and (A-T)$_N$ chains. 

\begin{figure}
\includegraphics[scale=0.82]{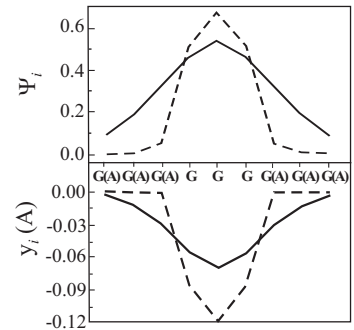}
\caption{The wave function $\Psi_i$ and the lattice 
displacement $y_i$ for the polarons formed in GGGGGGGGG 
(solid line) and AAAGGGAAA (dashed line) chains.} 
\label{fig:fig1}
\end{figure}

With the values of coupling constants derived for 
different base pairs we now study the properties of the
polaronic state in the poly and mixed DNA chains. 
In the poly(dG)-poly(dC) 
and the poly(dA)-poly(dT) DNA molecules, according to our 
calculations the polaron occupies mostly 7--9 sites. 
In the mixed DNA chain, the polaron stretching is limited 
by the difference between the coupling constants $\chi_i$, 
and on-site energies $\epsilon_i$, for A-T and 
G-C pairs (see Table~\ref{tab:table1}). The results for 
GGGGGGGGG and AAAGGGAAA chains are shown in Fig.~1. 
The polaron in the AAAGGGAAA structure is mostly localized 
within the GGG due to high potential barriers between 
guanines and adenines (1.08 eV) \cite{Saito}.

\begin{figure}
\includegraphics[scale=0.82]{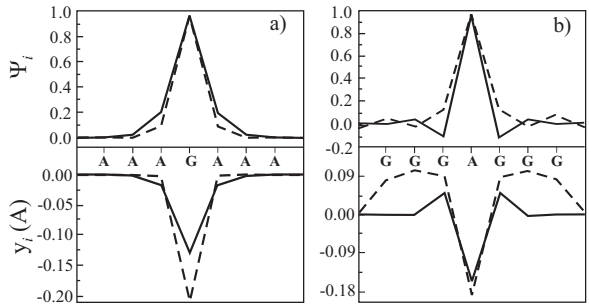}
\caption{The wave function $\Psi_i$ and the lattice 
displacement $y_i$ for polarons formed on 
(a) G-C for $\chi_i$=0.6 eV/\r{A} (solid line) 
and  $\chi_{\mathrm{G-C}}$=1.6 eV/\r{A}, 
$\chi_{\mathrm{A-T}}=0.4$ eV/\r{A} (dashed line) 
and on (b) A-T $\chi_i$=0.6 eV/\r{A} (solid line) 
and $\chi_{\mathrm{G-C}}$=0.9 eV/\r{A},  
$\chi_{\mathrm{A-T}}$=1.05 eV/\r{A} (dashed line).} 
\label{fig:fig2}
\end{figure}

The value of the coupling constant also determines 
the polaron stretching, but its effect is strong 
only in the structure with low potential barriers. An example 
of such structures is the AAAAGAAAA 
chain where the energy gap between the A-T and the G-C is only 
0.4 eV. The results for the AAAAGAAAA structure are shown in 
Fig.~\ref{fig:fig2} (a) for two cases: (i) when the value of 
coupling constant is the same over the whole chain and is equal 
to $\chi_i$=0.6 eV/\r{A} \cite{Saito}, and (ii) when the coupling 
constant is different for G-C and A-T base pairs. Clearly, the 
introduction of different coupling  constants $\chi_i$ for A-T and 
G-C pairs provides stronger localization of the polaron within
the G-C trap.

The effect of the coupling constant on the polaron localization in 
the GGGGAGGGG chain is illustrated in Fig.~\ref{fig:fig2} (b).
Actually, for this chain the polaron vibration 
mode is outside of the lattice band of the A-T site \cite{Bishop2}. 
When the coupling constant $\chi_i$=0.6 eV/\r{A}
is the same over the whole DNA chain, the vibration
mode is only marginally delocalized. The energy
of this state is $-0.81$ eV, while the potential barrier between 
(G-C)$_N$ and A-T site is $-$0.87 eV. The polaron in this 
case is almost localized at the A-T site. 
When we introduce the dependence of the coupling constant on the 
site type [Fig.~\ref{fig:fig2} (b) (dashed line)], the 
energy of the state becomes $-0.56$ eV and the
polaron becomes delocalized over three nearest
G-C sites from each side of the A-T pair.

To study polaron tunneling between the DNA traps, 
we first compare the energies of the polaronic 
states in different types of traps.  
Localization of the polaron in the (G-C)$_N$ traps shifts 
the on-site energy $\epsilon_i$ to a lower value. This 
is the energy of the polaron which is the eigenvalue of the 
Hamiltonian corresponding to Eq.~(\ref{eq:one}).
This energy can also be estimated from the site energy $\epsilon_i$ and 
the electronic energy $\chi_i y_i$ \cite{Bishop2}
\begin{equation}
E_{\rm tot}\simeq \sum_{i=1}^N\chi_i 
y_i+ \sum_{i=1}^N\epsilon_i/N=\frac12\lambda_N+\sum_{i=1}^{N}\epsilon_i/N.
\label{eq:eight}
\end{equation}
For the energy difference between the polaronic states in different 
traps we have found the values $-0.20$ eV for G-C and (G-C)$_2$ traps and 
$-0.43$ eV for G-C and (G-C)$_3$ traps. Inclusion of inner-sphere reorganization energy 
into the charge transfer model has brought down these values from $\Delta\epsilon=-0.47$ eV and
$\Delta\epsilon=-0.68$ eV, respectively \cite{Saito}. 
A direct comparison with the experimental results 
in the solvent \cite{lewisa} would require 
evalution of the solvent reorganization energy \cite{berash1}, 
which is beyond the scope of this work. However, for the
results that follow, in particular for the polaron migration dynamics, 
the solvent contribution perhaps
is not the dominant one.

The low energy gap between the states of the (G-C)$_N$ traps 
results in the competition between two processes in the 
mixed DNA \cite{Schuster2}: (i) the trapping 
of the polaron within the trap and (ii) 
the tunneling of the polaron between the (G-C)$_N$ traps. 
To study the problem of polaron tunneling between the DNA traps 
we have performed numerical simulation of the polaron dynamics 
in a mixed DNA chain. Here the first G-C trap 
is a donor with localized charge 
on it in the initial state and the (G-C)$_3$ trap is an acceptor. 
Since the system is initially in the nonequlibrium state the 
polaron will tunnel from the donor to the acceptor. 
For the system without any additional traps, i.e. a DNA chain 
with only the donor and the acceptor [(G)(A)$_n$(G)$_3$], 
we have found an exponential dependence of the 
tunneling rate on the tunneling distance.  
Let $t_n$ be the tunneling time for the structure (G)(A)$_n$(G)$_3$.
We then have for the normalized rate, $t_1/t_n = 0.6 \times 10^{-1}$ 
for $n=2$, $1.4\times10^{-3}$ for $n=3$, and $0.6 \times 10^{-5}$ 
for $n=4$ ($t_n=0.5$ ns). These data are in a good agreement with 
the tunneling rates in the experimental results \cite{giese}.

\begin{figure}
\includegraphics[scale=0.36]{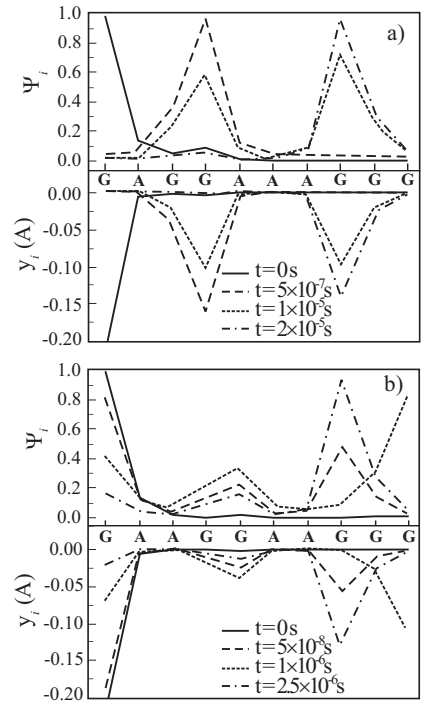}
\caption{The dynamics of propagation of the wave function 
$\Psi_i$ and the lattice displacement $y_i$ in the 
chains: (a) G(A)$_1$GG(A)$_3$GGG and (b) G(A)$_2$GG(A)$_2$GGG.} 
\label{fig:fig3}
\end{figure}

The new features in the polaron tunneling process 
is observed for the mixed DNA structure with additional 
trap between the donor and the acceptor. Here we study the 
system with the (G-C)$_2$ trap in the (A-T)$_6$ molecular 
bridge. The dispersion of the site energies of the G-C base pairs within the 
(G-C)$_3$ where the guanine at the end site has higher 
site energy than guanines located
close to the sequence center \cite{berash}, has been taken into account.
In the case of the (G-C)$_2$ trap located close to 
the donor site (G-C) the polaron is stretched over the donor 
and the trap. As a result the polaron quickly tunnels to 
the (G-C)$_2$ trap [Fig.~\ref{fig:fig3}(a)]. The polaron 
occupation process takes some time and finally the polaron tunnels to 
the acceptor. In this case the tunneling from the donor to 
the acceptor states has the sequential nature and the tunneling 
processes from the donor to the trap and from the trap to the acceptor 
are uncorrelated.

When the 
(G-C)$_2$ trap is placed exactly 
in the middle of the (A-T)$_{6}$ bridge, 
a significant change in the polaron tunneling dynamics is observed
[Fig.~\ref{fig:fig3} (b)]. In this case, the rate of charge 
tunneling from the donor to the trap is almost 
equal to the rate of tunneling from the trap to the acceptor. 
Therefore, the polaron is only partially localized on the trap 
and the final polaron tunneling from the donor to the acceptor is 
a coherent process. The curve for $t=10^{-6}$ sec in Fig.~\ref{fig:fig3} (b)
shows the occurence of the resonance effect due to the coincidence of the trap site energy
with the site energy of the last guanine within the (G-C)$_3$ acceptor.
The rate of charge tunneling in Fig.~\ref{fig:fig3} 
is in good agreement with the 
experimental results \cite{giese2}, where the transfer from a donor to 
an acceptor in similar systems was estimated to be $10^{-8}-10^{-6}$ sec. 

\begin{figure}
\includegraphics[scale=0.45]{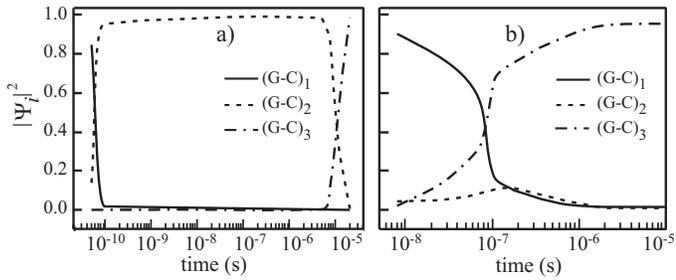}
\caption{The dependence of the occupation probability $|\Psi_i|^2$ 
on time for the chains: (a) G(A)$_1$GG(A)$_{3}$GGG and 
(b) G(A)$_2$GG(A)$_2$GGG.} 
\label{fig:fig4}
\end{figure}

In Fig.~\ref{fig:fig4} the dependence of the occupation 
probability of different traps within the DNA chain is 
shown for (a) G(A)$_1$GG(A)$_{3}$GGG and (b) 
G(A)$_2$GG(A)$_2$GGG structures. We again see a completely 
different nature of tunneling for different positions of 
the trap. When the trap is close to the donor, the charge 
transfer process is the sequential incoherent tunneling, i.e.,
when the polaron spends a long time within the trap. 
But if the trap is moved closer to the center of the 
tunneling bridge 
then the tunneling becomes coherent. 
However, when the position of the (G-C)$_2$ trap is closer 
to the acceptor as in the G(A)$_3$GG(A)$_{1}$GGG structure, 
the polaron is not localized on the trap
but tunnels directly from the donor to the acceptor.
Since the trap does not participate in the charge transfer, 
the width of the potential barrier for the polaron covers 
the whole molecular bridge (A)$_3$GG(A)$_1$ and coherent 
tunneling from the donor to the acceptor occurs in the range 
of $t=200$ sec. This is $10^{7}$ times slower than the time 
for coherent tunneling through the trap [see in Fig.3(b) and Fig.4(b)].
Therefore, the transfer mechanism for the G(A)$_3$GG(A)$_{1}$GGG 
sequence is similar to that for
the G(A)$_{6}$GGG structure.

In conclusion, from the results of the ab initio
quantum mechanical calculations, we obtained the 
charge-vibration coupling constants in the 
Peyrard-Bishop-Holstein model for polarons formed 
in the (G-C)$_N$ and the (A-T)$_N$ DNA molecules. 
We have found that the coupling constants are  
larger for the (G-C)$_N$ complex than for the (A-T)$_N$. From 
the calculated values of the coupling constants we have 
studied the energetics and the structure of the polaron in 
different DNA sequences. In the poly-DNA molecule, the polaron 
occupies nine DNA base pairs, while in the mixed DNA 
the size of the polaron is strongly affected by the 
potential gap between the A-T and the G-C sites. 
In addition to the properties of the 
stationary polaronic state, we have also studied the 
dynamics of the polaron tunneling from the donor to the acceptor.  
We have found a very strong 
dependence of the tunneling rates on the structure of the 
tunneling bridge. The position of additional traps within 
the bridge strongly affects the nature of the tunneling
process and the rates. By changing the position 
we can change the tunneling rate upto 
seven orders of magnitude. To have the fastest 
tunneling rate we need to have coherent tunneling, i.e. 
tunneling to each of the traps should be almost equal to the 
escape rate from the trap.
Our calculations are restricted to low temperatures. Therefore, 
the dynamics of the polaron in our study is completely due 
to quantum mechanical processes. 

This work has been supported by the Canada Research
Chair Program and a Canadian Foundation for Innovation 
(CFI) grant.


\begin{thebibliography}{99}
\bibitem[\ddag]{byline} Electronic mail:
tapash@physics.umanitoba.ca
\bibitem{review} {\it Long-range charge transfer in DNA}, 
edited by, G.B. Schuster (Springer-Verlag, Heidelberg, 
New York 2004).
\bibitem{ratner}
M.A. Ratner, and J. Jortner, {\it Molecular Electronics} 
(Blackwell, Oxford, 1997).
\bibitem{jortner}
J. Jortner, M. Bixon, T. Langenbacher, and M.E. Michel-Beyerle, 
Proc. Natl. Acad. Sci. (USA) {\bf 95}, 12759, (1998). 
\bibitem{berlin}
Y.A. Berlin, A.L. Burin, and M.A. Ranter, J. Phys. Chem. A 
{\bf 104}, 443 (2000).
\bibitem{conw}
E. Conwell, Top. Curr. Chem. {\it 237}, 73 (2004).
\bibitem{yoo}
K.-H. Yoo, et al., Phys. Rev. Lett. {\bf 87} 198102 (2001).
\bibitem{kasumov}
A.Y. Kasumov, et al. Science {\bf 291} 280 (2001).
\bibitem{kawai}
M. Taniguchi, and T. Kawai, Physica E {\bf 33}, 1 (2006).
\bibitem{porath}
D. Porath, A. Bezryadin, S. deVries, C. Dekker, Nature 
{\bf 403}, 635 (2000).
\bibitem{wang}
X.F. Wang and T. Chakraborty, Phys. Rev. Lett. {\bf 97}, 
106602 (2006)
\bibitem{macia}
S. Roche, E. Macia, Mod. Phys. Lett. B. {\bf 18} 847 (2004).
\bibitem{mark1}
R.A. Marcus, J. Chem. Phys. {\bf 24}, 966 (1956).
\bibitem{Bishop1}
P. Maniadis, et al. Phys. Rev. E {\bf 72}, 021912 (2005).
\bibitem{Bishop2}
P. Maniadis, et al. Phys. Rev. B. {\bf 68}, 174304 (2003).
\bibitem{berash}
J. Berashevich, and T. Chakraborty, cond-mat/0703764. 
\bibitem{Saito}
H. Sugiyama and I. Saito, J. Am. Chem. Soc. {\bf 118}, 
7063 (1996).
\bibitem{lewisa}
F.D. Lewis, et al. J. Am. Chem. Soc. {\bf 122} 12037 (2000).
\bibitem{berash1}
J. Berashevich, and T. Chakraborty, J. Chem. Phys. {\bf 126}, 035104 (2007).
\bibitem{Schuster2}
J. Joseph, G.B. Shuster, J. Am. Chem. Soc. {\bf 128}, 6070 (2006). 
\bibitem{giese}
F.D. Lewis, et al. Angew. Chem. Int. Ed. {\bf 45}, 7982 (2006).
\bibitem{giese2}
F.D. Lewis, et al. J. Am. Chem. Soc. {\bf 125} 4850 (2003).
\end{thebibliography}
\end{document}